\documentclass[twocolumn,showpacs,amsmath,amssymb]{revtex4}
\usepackage{amsmath}
\usepackage{longtable}
\usepackage{dcolumn}
\begin{document}
\title{Proton-neutron interaction near closed shells} 
\date{\today}
\author{A. Covello}
\author{L. Coraggio}
\author{A. Gargano}
\author{N. Itaco} 
\affiliation{Dipartimento di Scienze Fisiche, Universit\`a di Napoli
Federico II, and Istituto Nazionale di Fisica Nucleare, 
 Complesso Universitario di Monte S. Angelo, Via Cintia, I-80126 Napoli, Italy}

\begin{abstract}
Odd-odd nuclei around double shell closures are a direct source of information on the proton-neutron interaction between valence nucleons. We have performed shell-model calculations for doubly odd nuclei close to $^{208}$Pb, $^{132}$Sn and $^{100}$Sn using realistic effective interactions derived from the CD-Bonn nucleon-nucleon potential. The calculated results are compared with the available experimental data, attention being focused on particle-hole and particle-particle multiplets. While a good agreement is obtained for all the nuclei considered, a detailed analysis of the matrix elements of the effective interaction shows that a stronger core-polarization contribution seems to be needed in the particle-particle case.

\end{abstract}

\maketitle

\section{Introduction}

Odd-odd nuclei around double shell closures provide the best testing ground for the matrix elements of the proton-neutron interaction between valence nucleons. In this context, of special interest are nuclei in the close vicinity to $^{208}$Pb, $^{132}$Sn and $^{100}$Sn, which show strong shell closures for both protons and neutrons. From the experimental point of view, the $^{208}$Pb neighbors have been extensively investigated and a large amount of data is available for them. On the other hand, the $^{132}$Sn and $^{100}$Sn neighbors lie well away from the valley of stability, which makes it very difficult to obtain information on their spectroscopic properties. In recent years, however, substantial progress in the development of spectroscopic techniques has opened the way to the exploration of nuclear structure in the regions of shell closures off stability. This has led to new experimental data for the immediate odd-odd neighbors of both $^{132}$Sn and $^{100}$Sn. The information presently available for nuclei of this kind, although still rather scanty, is of great value for the understanding of the effective proton-neutron interaction in these two regions as well as for a comparison with the proton-neutron interaction in the $^{208}$Pb region.

During the past several years, we have studied a number of nuclei around $^{208}$Pb,  $^{132}$Sn and $^{100}$Sn within the framework of the shell model employing realistic effective interactions derived from modern nucleon-nucleon ($NN$) potentials. In most of these studies, however, we have been concerned with nuclei having few identical valence particles or holes. A summary of the results obtained through 2000 is given in Ref. 
\cite{cov1}.

More recently \cite{cor02a,cov03a}, we have turned our attention to nuclei with unlike valence nucleons to try to gain information on the neutron-proton effective interaction. 
The main aim of this paper is to report on some selected results of our current work along these lines, which we have obtained starting from the CD-Bonn free $NN$ potential 
\cite{machl01}. In particular, we consider the six odd-odd nuclei 
$^{208}$Bi, $^{210}$Bi, $^{132}$Sb, $^{134}$Sb, $^{98}$Ag and $^{102}$In, and focus
attention on the particle-hole and particle-particle multiplets, which offer the opportunity to test directly the matrix elements of our calculated realistic effective interaction.

To place this study in its proper perspective, it should be mentioned that the particle-hole multiplets in the  $^{208}$Pb region were the subject of great theoretical and experimental interest \cite{ersk64,alf68,kuo68,moin69} some thirty-five years ago. In particular, the spectrum of $^{208}$Bi was extensively studied through pick-up and stripping reactions and several proton-neutron hole multiplets were identified \cite{ersk64,alf68}. The pattern exhibited by the observed multiplets was well reproduced by the shell-model calculations of Ref. \cite{kuo68}, where particle-hole matrix elements derived from the Hamada-Johnston $NN$ potential \cite{ham62} were used. 

Despite these early achievements in the understanding of the effective proton-neutron interaction around closed shells, little work along the same lines has been done during the last three decades. The new data which are becoming available in the $^{132}$Sn and $^{100}$Sn regions, as well as the prospect of spectroscopic studies of unstable nuclei opened up by the development of radioactive ion beams, has motivated us to undertake the present study and perform shell-model calculations making use of a modern $NN$ potential and improved many-body methods for deriving the effective interaction.

The outline of the paper is as follows. In Sec. II we give a brief description of our calculations. Our results are presented and compared with the experimental data in Sec. III, where we also discuss the effects of the core polarization. In Sec. IV we draw the conclusions of our study.

\section{Calculations}

In our calculations for the Bi isotopes we assume that $^{208}$Pb is a closed core and let the valence proton and neutron hole occupy the six single-particle levels $0h_{9/2}$, $1f_{7/2}$, $0i_{13/2}$, $1f_{5/2}$, $2p_{3/2}$ and $2p_{1/2}$ of the 82-126 shell, while for the valence neutron the model space includes the seven orbits $1g_{9/2}$, $0i_{11/2}$, $0j_{15/2}$, $2d_{5/2}$, $3s_{1/2}$, $1g_{7/2}$, and $2d_{3/2}$. Similarly, for the Sb isotopes we assume that the valence proton and neutron hole occupy the five single-particle levels $0g_{7/2}$, $1d_{5/2}$, $1d_{3/2}$, $2s_{1/2}$ and $0h_{11/2}$ of the 50-82 shell, while the valence neutron outside the $^{132}$Sn core occupies the orbits of the 82-126 shell. As regards $^{98}$Ag and $^{102}$In, we assume that the valence neutrons outside the $^{100}$Sn core occupy the orbits of the 50-82 shell, while for the proton holes the model space includes the four orbits $0g_{9/2}$, $1p_{1/2}$, $1p_{3/2}$ and $0f_{5/2}$ of the 28-50 shell. 

As regards the choice of the single-particle and single-hole energies, we have proceeded as follows. For the Bi isotopes we have taken them from the experimental spectra \cite{nndc} of $^{209}$Bi, $^{209}$Pb and $^{207}$Pb. In the same way, for the Sb isotopes we have made use of the experimental spectra of $^{133}$Sb \cite{sanch99}, $^{133}$Sn \cite{hoff96} and $^{131}$Sn \cite{nndc}. In the spectra of the two former nuclei, however, some single-particle levels are still missing. More precisely, this is the case of the proton $2s_{1/2}$ and neutron $0i_{13/2}$ levels, whose energies have been taken from Refs. \cite{andr97} and \cite{cor02b}, respectively, to which we refer for details. As regards the neutron hole energy $\epsilon_{h_{11/2}}$ in $^{131}$Sn, we have adopted the value of 0.100 MeV which has been recently suggested in Ref. \cite{gen00}. This is somewhat smaller than that reported in Ref. \cite{nndc}. For $^{98}$Ag  and $^{102}$In the single-particle and single-hole energies cannot be taken from experiment, since no spectroscopic data are yet available for 
$^{101}$Sn and $^{99}$In. Therefore, we have taken them from Ref. \cite{andr96} and Ref. \cite{cor00}, where they have been determined by an analysis of the low-energy spectra of the Sn isotopes with $A \leq 111$ and of the N=50 isotones with $A \geq 89$. 
For completeness, our adopted values for the single-particle and single-hole energies are reported in Tables I-III.

As already mentioned in  the Introduction, in our shell-model calculations we have made use of a realistic effective interaction derived from the CD-Bonn free nucleon-nucleon potential \cite{machl01}. This high-quality $NN$ potential, which is based upon meson exchange, fits very accurately ($\chi^2$/datum $\approx 1$) the world $NN$ data below 350 MeV available in the year 2000.

The shell-model effective interaction $V_{\rm eff}$ is defined, as usual, in the following way. In principle, one should solve a nuclear many-body Schr\"odinger equation of the form 
\begin{equation}
H\Psi_i=E_i\Psi_i, 
\end{equation}
with $H=T+V_{NN}$, where $T$ denotes the kinetic energy. This full-space many-body problem is reduced to a smaller model-space problem of the form
\begin{equation}
PH_{\rm eff}P \Psi_i= P(H_{0}+V_{\rm eff})P \Psi_i=E_iP \Psi_i.
\end{equation}
Here $H_0=T+U$ is the unperturbed Hamiltonian, $U$ being an auxiliary potential introduced to define a convenient single-particle basis, and $P$ denotes the projection operator onto the chosen model space,
\begin{equation} P = \sum_{i=1}^d |\psi_i\rangle \langle \psi_i|,
\end{equation}
$d$ being the dimension of the model space and $|\psi_i\rangle$ the eigenfunctions of $H_0$. The effective interaction $V_{\rm eff}$ operates only within the model space $P$. In operator form it can be schematically written \cite{kuo80} as

\begin{equation}
V_{\rm eff} = \hat{Q} - \hat{Q'} \int \hat{Q} + \hat{Q'} \int \hat{Q} \int
\hat{Q} - \hat{Q'} \int \hat{Q} \int \hat{Q} \int \hat{Q} + ~...~~,
\end{equation}

\noindent where $\hat{Q}$, usually referred to as the $\hat{Q}$-box, is a vertex function composed of irreducible linked diagrams, and the integral sign represents a
generalized folding operation. $\hat{Q'}$ is obtained from $\hat{Q}$ by removing terms of first order in the interaction. Once the $\hat{Q}$-box is calculated,
the folded-diagram series of eq. (4) can be summed up to all orders by iteration methods, as, for instance, the Lee-Suzuki one \cite{lee80,suz80}. 

A main difficulty encountered in the derivation of $V_{\rm eff}$ from any modern $NN$ potential  is the existence of a strong repulsive core which prevents its direct use in nuclear structure calculations. This difficulty is usually overcome by resorting to the well known Brueckner $G$-matrix method. Here, we have made use of a new approach \cite{bogn02} which has proved to be an advantageous alternative \cite{bogn02,cov02,cor03} to the use of the above method. The basic idea underlying this approach is to construct a low-momentum $NN$ potential, $V_{\rm low-k}$, that preserves the physics of the original potential $V_{NN}$ up to a certain cut-off momentum $\Lambda$. In particular, the scattering phase shifts and deuteron binding energy calculated by $V_{NN}$ are reproduced by $V_{\rm low-k}$. The latter is a smooth potential that can be used directly as input for the calculation of shell-model effective interactions. A detailed description of our derivation of $V_{\rm low-k}$ can be found in Ref. \cite{bogn02}, where a criterion for the choice of the cut-off parameter $\Lambda$ is also given. We have used here the value $\Lambda=2.1$ fm$^{-1}$. 

Once the $V_{\rm low-k}$ is obtained, the calculation of the matrix elements of the effective interaction is carried out within the framework of the folded-diagram method outlined above. The key feature of this approach is that there is no need for $G$-matrix procedures to eliminate the effects of a strongly repulsive core, since the  $V_{\rm low-k}$ can be used directly in the calculation of the vertex function $\hat{Q}$-box. 
The calculation of $\hat{Q}$, which is in principle an infinite sum of irreducible diagrams, can only be made approximately by selecting certain classes of diagrams. In our calculations 
we have included in $\hat{Q}$ all the 1-body  and 2-body diagrams up to second order in 
$V_{\rm low-k}$. For instance, for the particle-particle case these are precisely the diagrams shown in Ref. \cite{jian92}. In this connection, it should be pointed out that the proton-neutron matrix elements for $^{208}$Bi, $^{132}$Sb, $^{98}$Ag and $^{102}$In have been explicitly derived in the particle-hole formalism. A description of the derivation of the particle-hole effective interaction is given in Ref. \cite{cor02a}.

To summarize, there are three main steps in our derivation of $V_{\rm eff}$. We first derive the low-momentum  $V_{\rm low-k}$ from the CD-Bonn potential, then calculate the $\hat{Q}$-box including diagrams up to second order in $V_{\rm low-k}$, and finally obtain $V_{\rm eff}$ by summing up the folded diagram series (4) by means of the Lee-Suzuki iteration method.

\section{Results and comparison with experiment}

In this section, the results of our calculations for the proton-neutron multiplets around $^{208}$Pb, $^{132}$Sn, and $^{100}$Sn are presented and compared with the experimental data. In the first two mass regions, both proton-neutron hole and proton-neutron
multiplets are considered by studying $^{208,210}$Bi and $^{132,134}$Sb.
These nuclei, all having only two valence nucleons, are the most appropriate ones to study the effects of the proton-neutron interaction. 
As regards doubly magic $^{100}$Sn, the two immediate odd-odd neighbors for which some experimental information is presently available are $^{98}$Ag and $^{102}$In, both with four valence nucleons. We have therefore considered these two nuclei to study the proton hole-neutron multiplets in this region. We present our results  separately for each mass region in the three following subsections. All calculations have been performed using the OXBASH shell-model code \cite{OXBASH}.

\subsection{$^{208}$Pb region: $^{208}$Bi and $^{210}$Bi}

The particle-hole multiplets in $^{208}$Bi were long ago found to exhibit a peculiar behavior \cite{ersk64,alf68}, namely the states with the minimum and maximum $J$ 
have the highest excitation energy, while the state with next to highest $J$ is the lowest one. As mentioned in the Introduction, the results of the shell-model calculations of Ref. \cite{kuo68} turned out to be consistent with this pattern.

In Fig. 1 our results for three multiplets are reported and compared with the experimental data \cite{nndc}. Note that the energies are relative to the $5^+$ ground state, which is a member of the doublet $\pi h_{9/2} \nu p_{1/2}^{-1}$. We see that the calculated energies account for the pattern of the experimental multiplets, the quantitative agreement being remarkably good with discrepancies  well below 100 keV for almost all the states.
It should be pointed out that our calculations give a better description of the experimental data than the early realistic calculations of Ref. \cite{kuo68}.

Let us now come to $^{210}$Bi. In Fig. 2 the calculated $\pi h_{9/2} \nu g_{9/2}$ multiplet (represented by  open circles) is shown and compared with the 
experimental data \cite{nndc}. This multiplet shows a breakdown of the Nordheim strong rule 
\cite{nord50} in that the $1^-$ state is the ground state while the $0^-$ one lies at about 50 keV excitation energy. We see that our calculation predicts for the ground state $J^{\pi}=0^{-}$ with the $1^-$ state at about 150 keV excitation energy.
This is, however, the most significant discrepancy, all other experimental excitation energies being reproduced within few tens  of  keV. As a whole, the pattern of this multiplet as compared to that of the particle-hole multiplets is still parabolic, but concave downward. Of course this stems from the fact that the interaction changes from repulsive to attractive. 

It should be pointed out that the behavior of all the multiplets considered above is directly related to the proton-neutron effective interaction since their members are of almost pure configuration. In this context, it is quite interesting to study the renormalization of the effective interaction, which we have taken into account through second-order diagrams in $V_{\rm low-k}$. By way of illustration, we report in Tables IV and V the diagonal matrix elements of the effective interaction for the particle-hole $\pi h_{9/2} \nu f_{5/2}^{-1}$ and particle-particle $\pi h_{9/2} \nu g_{9/2}$ configuration, respectively.  In both tables we also show the corresponding matrix elements 
of $V_{\rm low-k}$ and $<V_{\rm ph}>$, the latter representing  the contribution  of the particle-hole core-polarization diagram, the so-called ``bubble". Here, we do not consider the contributions coming from either other second-order diagrams or folded diagrams, both of which we have 
found irrelevant for the present discussion. From Table IV, we see that the 
matrix elements of $V_{\rm ph}$, although not very large in magnitude, are quite relevant in  determining the pattern of the considered particle-hole multiplet. In fact, they are repulsive for states with minimum and maximum $J$ and attractive for $J=J_{\rm max}-1$. It is worth noting that the repulsive contribution to the $J_{\rm max}$ state is essential to place  it above the state with  $J=J_{\rm min}+1$. For the particle-particle case, Table V shows that the core-polarization contribution is essentially repulsive, except for  the $J^{\pi}=1^{-}$ state. This goes in the right direction, as it reduces the spacing between the $1^{-}$ and $0^{-}$ states. However, as is evident from Fig. 2, it is not sufficient to produce the inversion of these two states.

On these grounds, we have found it interesting to modify our effective interaction 
for $^{210}$Bi by simply increasing the diagonal matrix elements of $V_{\rm ph}$ for the $\pi h_{9/2} \nu g_{9/2}$ configuration. It turns out that all the calculated energies go in the right direction, a factor of 2.6 being sufficient to give the correct ground state.  The results of this calculation (represented by open squares) are shown in Fig. 2. It can be seen that they practically overlap the experimental ones.

\subsection{$^{132}$Sn region: $^{132}$Sb and $^{134}$Sb}

The particle-hole and particle-particle nuclei $^{132}$Sb and $^{134}$Sb play in the $^{132}$Sn region the role played by $^{208}$Bi and $^{210}$Bi in the {$^{208}$Pb region.
In this connection, it is worth mentioning  that the resemblance between the spectroscopy of these two regions has been emphasized in several recent papers \cite{forn01}.

The nucleus $^{132}$Sb  has been the subject of our study of Ref. \cite{cor02a}, to which we refer for details. Some multiplets have been also reported in Ref. \cite{garg03} together with some results for $^{134}$Sb. For completeness, we report in Fig. 3 the 
$\pi g_{7/2} \nu d_{3/2}^{-1}$ and 
$\pi g_{7/2} \nu h_{11/2}^{-1}$ multiplets in  $^{132}$Sb while Fig. 4 shows our results for the $\pi g_{7/2} \nu f_{7/2}$ multiplet in $^{134}$Sb. Note that these multiplets are just the counterparts of those considered above for $^{208}$Bi and $^{210}$Bi (aside from the doublet $\pi g_{9/2} \nu p_{1/2}^{-1}$).
Comments on the behavior of the multiplets as well as on the quality of agreement between theory and experiment can be found in Refs. \cite{cor02a,garg03}. In the latter paper, the diagonal matrix elements of $V_{\rm eff}$, $V_{\rm low-k}$, and $V_{\rm ph}$ for the 
$\pi g_{7/2} \nu d_{3/2}^{-1}$ and $\pi g_{7/2} \nu f_{7/2}$ configurations are reported. We emphasize here that the particle-hole and particle-particle matrix elements of the effective interaction in the $^{132}$Sn region show the same features as those evidenced above for the $^{208}$Pb region. In particular, we have found that also in this case the core-polarization  plays a significant role, but it is not sufficient to give the right spacing between the $0^{-}$ and $1^{-}$ states in the particle-particle multiplet of $^{134}$Sb (see Fig. 4).  We have then modified the effective interaction by multiplying the diagonal matrix 
elements of $V_{\rm ph}$ for the $\pi g_{7/2} \nu f_{7/2}$ configuration by the same factor used for $^{210}$Bi. The corresponding results are represented by open squares in Fig. 4, where we see that they are in excellent agreement with experiment.

\subsection{$^{100}$Sn region: $^{98}$Ag and $^{102}$In}

In the $^{100}$Sn region the counterpart of $^{208}$Bi and  $^{132}$Sb is $^{100}$In, which until now has not been accessible to spectroscopic studies.
We have therefore considered the two neighboring odd-odd isotopes  $^{98}$Ag and $^{102}$In , focusing attention on the $\pi g_{9/2}^{-1}\nu d_{5/2}$ multiplet for which some experimental information is available. The members of the calculated multiplet in both nuclei have been identified as those dominated by this configuration with the two remaining valence nucleons forming a zero-coupled pair.
In Figs. 5 and 6 we report the results of our calculations for  $^{98}$Ag and $^{102}$In, respectively, and compare them with the experimental data \cite{nndc}. We see that the agreement between experiment and theory is of the same quality as that obtained in the $^{208}$Pb and $^{132}$Sn regions, the largest discrepancy being 130 keV for the $5^+$ state in $^{98}$Ag.
The pattern of the calculated multiplets turns out to be similar to that of the multiplets in $^{208}$Bi and $^{132}$Sb. However, we find that the percentage of components other than those we have considered to characterize the multiplet is rather large. For instance, it reaches about 60\% for the  $7^+$ state in $^{102}$In}. A detailed discussion of the structure of the calculated states will be included in  a forthcoming publication.

The diagonal matrix elements of the effective interaction for the proton hole-neutron particle $\pi g_{9/2}^{-1}\nu d_{5/2}$ configuration are reported in Table VI, where we see that their behavior is quite similar to that occurring in the heavier-mass regions. 

\section{Summary and conclusions}

We have presented here some results of a shell-model study of odd-odd nuclei in the close vicinity to doubly magic $^{208}$Pb, $^{132}$Sn and $^{100}$Sn, where use has been made of  realistic effective interactions derived from the CD-Bonn $NN$ potential.

The main aim of this work has been to gain insight into the effects of the proton-neutron interaction on the pattern of the particle-hole and particle-particle multiplets. To this end, we have made a detailed comparison between the calculated results and the available experimental data as well as an analysis of the proton-neutron matrix elements of our effective interaction. It has turned out that the core polarization is essential for the effective interaction to produce a good description of all considered nuclei. In this context, we have found out that some discrepancies for the particle-particle multiplets
in both $^{210}$Bi and $^{134}$Sb are removed when the core polarization contribution  
is increased by a factor of 2.6. This is likely to be traced to the coupling of the single-particle motion to the octupole excitation in $^{208}$Pb and $^{132}$Sn.

A relevant outcome of our calculations is that for all the particle-hole multiplets the $J_{\rm min}$ and $J_{\rm max}$ states have the highest excitation energy, while the state with next to the highest $J_{\rm max}$ is the lowest, in agreement with the predictions of the Brennan-Bernstein coupling rule \cite{bren60}. 

In summary, we may conclude that the results of the present study show that realistic shell-model calculations are able to describe with quantitative accuracy the effects of the proton-neutron interaction around closed shells, which gives confidence in their predictive power. This is of particular value for nuclei in the $^{132}$Sn and $^{100}$Sn regions,
for which it is of utmost importance to gain more experimental information.

\begin{acknowledgments}
This work was supported in part by the Italian Ministero dell'Istruzione, dell'Universit\`a
e della Ricerca (MIUR). 
\end{acknowledgments}

\clearpage
\begin{figure}
\caption{Proton particle-neutron hole multiplets in $^{208}$Bi. The calculated results are represented by open circles while the experimental data by solid triangles. The lines are drawn to connect the points.}
\end{figure}

\begin{figure}
\caption{Proton-neutron  $\pi h_{9/2} \nu g_{9/2}$ multiplet in $^{210}$Bi. The calculated results are represented by open circles and squares (see text for comments) while the experimental data by solid triangles. The lines are drawn to connect the points.}
\end{figure}
\begin{figure}
\caption{Same as Fig. 1, but for $^{132}$Sb.}
\end{figure}
\begin{figure}
\caption{Same as Fig. 2, but for the  $\pi g_{7/2} \nu f_{7/2}$  multiplet in $^{134}$Sb.}
\end{figure}
\begin{figure}
\caption{Proton hole-neutron particle $\pi g_{9/2}^{-1}\nu d_{5/2}$ multiplet in $^{98}$Ag.
The conventions of the presentation are the same as those used in Fig. 1.}
\end{figure}
\begin {figure}
\caption{Same as Fig. 5, but for $^{102}$In.}
\end{figure}
\clearpage

\begin{table}

\caption{Proton single-particle and neutron single-hole and -particle energies (in MeV) for the $^{208}$Pb region.}
\begin{ruledtabular}
\begin{tabular}{c|c|c|c|c|c} 
$\pi (n,l,j)$&$\epsilon$&$\nu (n,l,j)^{-1}$&$\epsilon$&$\nu (n,l,j)$&$\epsilon$\\
\colrule
$0h_{9/2}$ & 0    & $2p_{1/2}$ & 0 & $1g_{9/2}$ & 0 \\
$1f_{7/2}$ & 0.896   & $1f_{5/2}$ & 0.570 &  $0i_{11/2}$ & 0.779 \\
$0i_{13/2}$ & 1.609   & $2p_{3/2}$ & 0.898 &  $0j_{15/2}$ & 1.423 \\
$1f_{5/2}$ & 2.826  & $0i_{13/2}$ & 1.633  &  $2d_{5/2}$ & 1.567 \\
$2p_{3/2}$ & 3.119   & $1f_{7/2}$ & 2.340 &  $3s_{1/2}$ & 2.032 \\
$2p_{1/2}$ & 3.633   &$0h_{9/2}$  &3.414 &   $1g_{7/2}$ & 2.491  \\
& & &  &   $2d_{3/2}$ & 2.538  \\
\end{tabular}
\end{ruledtabular}
\end{table}

\begin{table}
\caption{Proton single-particle and neutron single-hole and -particle energies (in MeV) for the $^{132}$Sn region.}
\begin{ruledtabular}
\begin{tabular}{c|c|c|c|c|c}
$\pi (n,l,j)$&$\epsilon$&$\nu (n,l,j)^{-1}$&$\epsilon$&$\nu (n,l,j)$&$\epsilon$\\
\colrule
$0g_{7/2}$ & 0    & $1d_{3/2}$ & 0 & $1f_{7/2}$ & 0 \\
$1d_{5/2}$ & 0.962   & $0h_{11/2}$ & 0.100 &  $2p_{3/2}$ & 0.854 \\
$1d_{3/2}$ & 2.439   & $2s_{1/2}$ & 0.332 &  $0h_{9/2}$ & 1.561 \\
$0h_{11/2}$ & 2.793  & $1d_{5/2}$ & 1.655  &  $2p_{1/2}$ & 1.656 \\
$2s_{1/2}$ & 2.800   & $0g_{7/2}$ & 2.434 &  $1f_{5/2}$ & 2.055 \\
&   &  & &   $0i_{13/2}$ & 2.694  \\
\end{tabular}
\end{ruledtabular}
\end{table}

\begin{table}
\caption{Proton single-hole and neutron single-particle energies (in MeV) for the $^{100}$Sn region.}
\begin{ruledtabular}
\begin{tabular}{c|c|c|c}
$\pi(n,l,j)^{-1}$&$\epsilon$ & $\nu (n,l,j)$& $\epsilon$\\
\colrule
$0g_{9/2}$ & 0    & $1d_{5/2}$ & 0 \\
$1p_{1/2}$ & 0.700   & $0g_{7/2}$ & 0.200 \\
$1p_{3/2}$ & 2.100   & $2s_{1/2}$ & 2.200 \\
$0f_{5/2}$ & 3.100  & $1d_{3/2}$ & 2.300  \\
&   &                $0h_{11/2}$ & 2.700  \\
\end{tabular}
\end{ruledtabular}
\end{table}

\begin{table}
\caption{Diagonal matrix elements of $V_{\rm eff}$, $V_{\rm low-k}$, and $V_{\rm ph}$ (in MeV) for the $\pi h_{9/2} \nu f_{5/2}^{-1}$
configuration in  $^{208}$Bi.}
\begin{ruledtabular}
\begin{tabular}{r|c|c|c}
 $J$ & $<V_{\rm eff}>$ & $<V_{\rm low-k}>$
& $<V_{\rm ph}>$\\
\colrule
2 & 0.479 & 0.512 & 0.102 \\
3 & 0.140 & 0.194& -0.008 \\
4 & 0.069 & 0.129 & -0.027 \\
5& 0.071  & 0.091& -0.001 \\
6& -0.020  & 0.044& -0.055 \\
7& 0.151  & 0.082& 0.087 \\
\end{tabular}
\end{ruledtabular}
\end{table}

\begin{table}
\caption{Diagonal matrix elements of $V_{\rm eff}$, $V_{\rm low-k}$, and $V_{\rm ph}$ (in MeV) for the $\pi h_{9/2} \nu g_{9/2}$ 
configuration in $^{210}$Bi.}
\begin{ruledtabular}
\begin{tabular}{r|c|c|c}
$J$ & $<V_{\rm eff}>$ & $<V_{\rm low-k}>$
& $<V_{\rm ph}>$\\
\colrule
0 & -0.524 &-0.514 & 0.020 \\
1 & -0.426 & -0.333& -0.093 \\
2 & -0.250 & -0.306 & 0.053 \\
3& -0.179  & -0.174& -0.002 \\
4& -0.089  & -0.153& 0.065 \\
5& -0.117  & -0.138& 0.025 \\
6& -0.027  & -0.084& 0.061 \\
7& -0.126  & -0.157& 0.034 \\
8& 0.015  & -0.038& 0.059 \\
9& -0.275  & -0.286& -0.003 \\
\end{tabular}
\end{ruledtabular}
\end{table}

\begin{table}
\caption{Diagonal matrix elements of $V_{\rm eff}$, $V_{\rm low-k}$, and $V_{\rm ph}$ (in MeV) for the $\pi g_{9/2}^{-1} \nu d_{5/2}$
configuration in $^{100}$In.}
\begin{ruledtabular}
\begin{tabular}{r|c|c|c}
$J$ & $<V_{\rm eff}>$ & $<V_{\rm low-k}>$
& $<V_{\rm ph}>$\\
\colrule
2 & 0.994 & 0.891 & 0.197 \\
3 & 0.328 & 0.335& 0.020 \\
4 & 0.171 & 0.249 & -0.038 \\
5& 0.196  & 0.175& 0.021 \\
6& -0.007  & 0.112& -0.102 \\
7& 0.487  & 0.205& 0.244 \\
\end{tabular}
\end{ruledtabular}
\end{table}

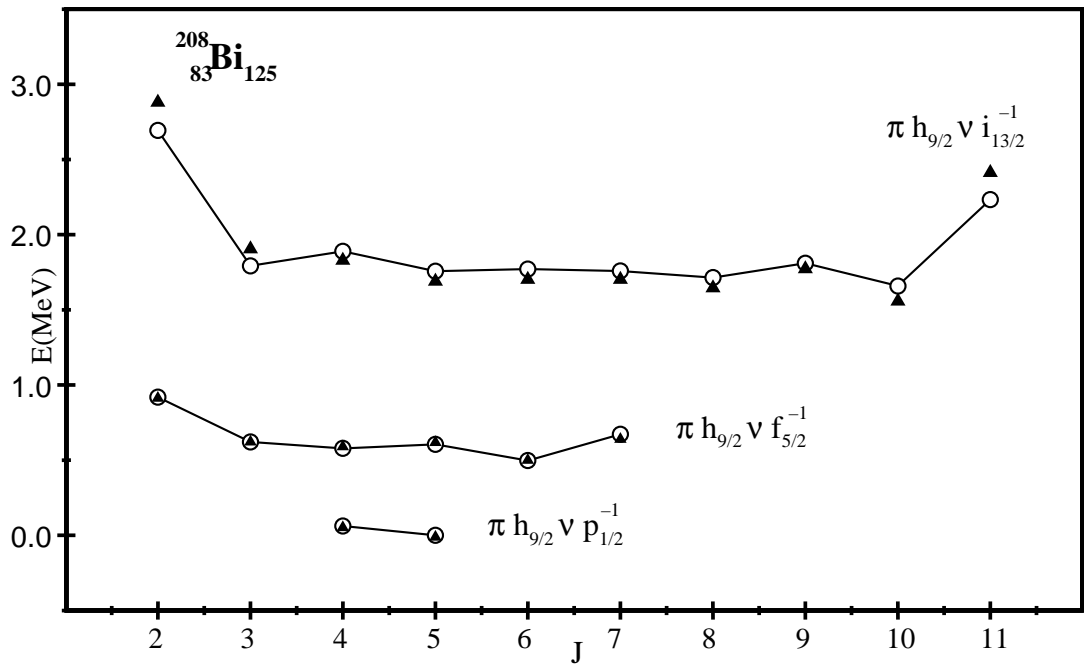
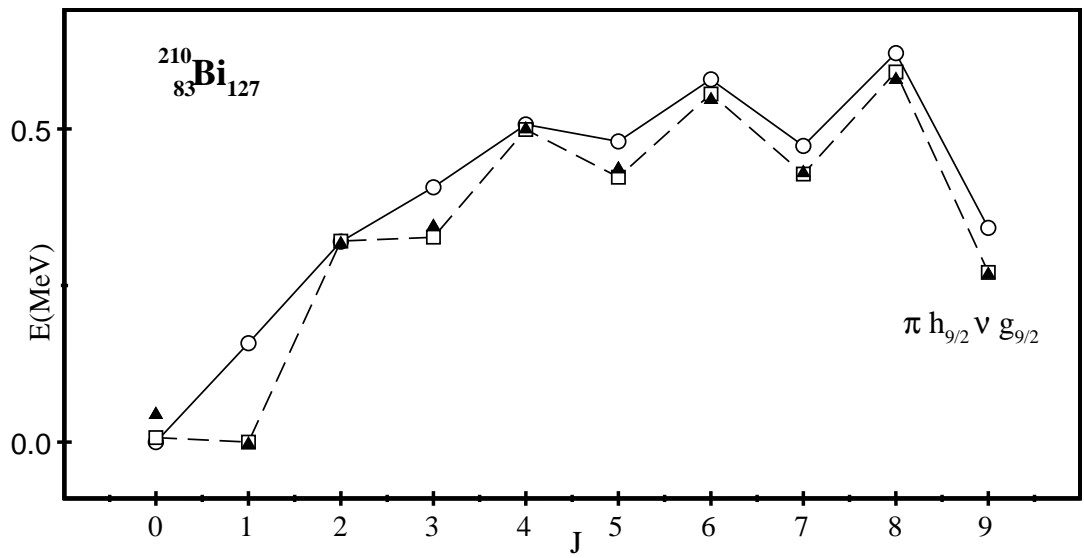
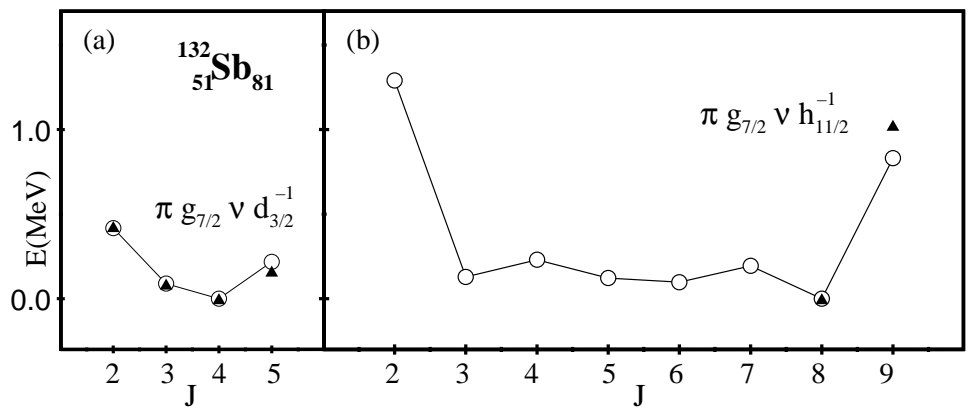
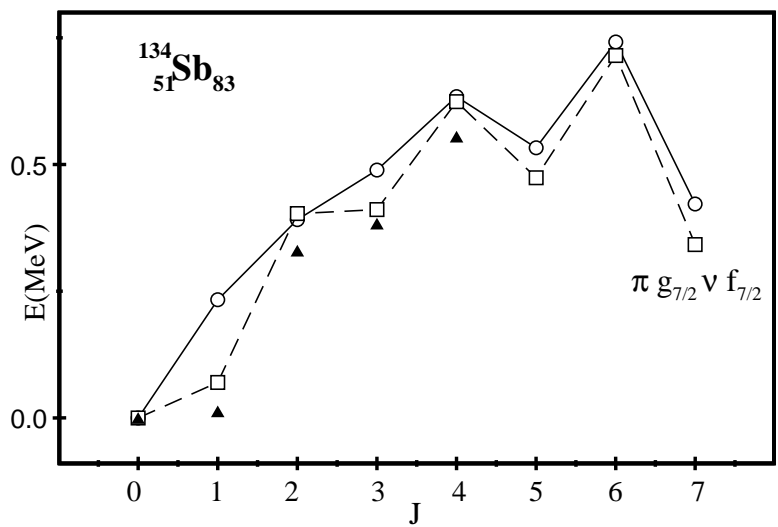
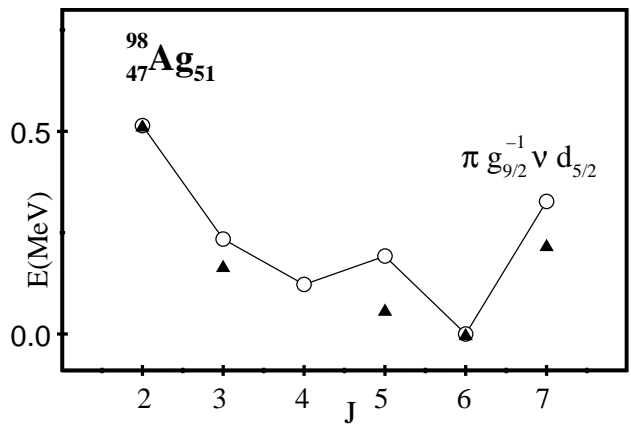
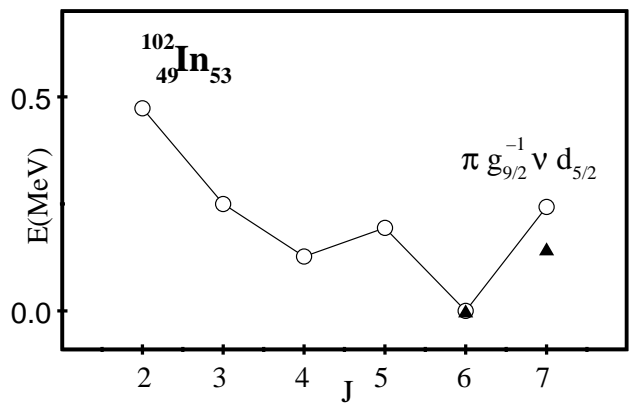
\end{document}